\documentclass[
 amsmath,amssymb,
preprint,
]{revtex4-1}

\usepackage{graphicx}% Include figure files
\usepackage{dcolumn}% Align table columns on decimal point
\usepackage{bm}% bold math
\usepackage{textcomp}
\usepackage{gensymb}
\usepackage{color,soul}

\usepackage[utf8]{inputenc}

\usepackage[version=4]{mhchem}
\usepackage{hyperref}
\begin{document}

\preprint{APLMat/MOSULLIVSIO}

\title{Interface doping and the effect of strain and oxygen stoichiometry on the transport and electronic structure properties of SrIrO\texorpdfstring{\textsubscript{3}}{3} heterostructures.}

\author{Wesley Surta}
%\email{w.surta@liverpool.ac.uk}
\affiliation{Department of Chemistry, University of Liverpool, Oxford Street, Liverpool, L69 7ZE, United Kingdom}%

\author{Saeed Almalki}
\affiliation{Department of Electrical Engineering and Electronics, University of Liverpool, Oxford Street, Liverpool, L69 7ZE, United Kingdom}%

\author{Ya-Xun Lin}
\affiliation{Department of Electrical Engineering and Electronics, University of Liverpool, Oxford Street, Liverpool, L69 7ZE, United Kingdom}%

%\author{Ivona Mitrovic}
%\affiliation{Department of Electrical Engineering and Electronics, University of Liverpool, Oxford Street, Liverpool, L69 7ZE, United Kingdom}%

\author{Tim Veal}
%\email{timveal@liverpool.ac.uk}
\affiliation{Department of Physics, University of Liverpool, Oxford Street, Liverpool, L69 7ZE, United Kingdom}%

\author{Marita O'Sullivan}
\email{mosulliv@liverpool.ac.uk}
\affiliation{Department of Physics, University of Liverpool, Oxford Street, Liverpool, L69 7ZE, United Kingdom}%

\date{\today}% It is always \today, today,

\begin{abstract}
The 5$d$ series semimetallic Dirac nodal-line perovskite \ce{SrIrO3} presents a promising system to study the interplay between spin-orbit coupling and electron-electron interactions in the epitaxial thin film geometry. The competition between the band splitting and the suppression of the correlation energy, from the extended d orbitals can influence the electronic and magnetic structure and lead to exotic quantum phases in such systems. This iridate member possesses a phase diagram which is acutely sensitive to both heterostructure growth conditions and geometric strain, which stabilises the orthorhombic phase over the monoclinic bulk crystal structure. The growth mode and lattice constants of this iridate perovskite have been observed to vary sharply with oxygen partial pressure and thickness during growth by pulsed laser deposition. In this study the structural and transport %and optical
properties originating from two sets of growth parameters have been compared, indicating distinct differences in the electronic structure at the Fermi edge. Density functional theory calculations of the convolved density of states compared with the X-ray photoemission spectra near the Fermi edge allow the identification of the states contributing to the electronic structure in this region. The spectral differences between these case studies indicate that controlling the growth regime could enable the tuning of the electronic structure to promote enhanced concentrations of one carrier type over another at the Fermi level in this nominally semimetallic system %while simultaneously allowing the transmission of visible light
. The control of the correlation-induced electronic transport parameters in this system will enable access to the spin-orbital entangled states and the topological crystalline properties associated with the honeycomb lattice in this Mott system.
\end{abstract}

\maketitle

\section{Introduction}
Materials possessing strong spin orbit coupling have excited interest in the quantum materials research community for their potential to exhibit predicted topological surface states \cite{Pesin2010376}, spin Hall effect \cite{EomPNAS2019} and superconductivity \cite{HYKee_Sr2IrO4, MacKensie2003}. The weak conductivity of iridium oxides is governed by the interplay between the spin-orbit and correlated electron interactions of the Ir$^{4+}$ 5$d$$^{5}$ electrons which has led to the study of Ruddlesden-Popper series of Sr$_{n+1}$Ir$_{n}$O$_{3n+1}$ (n = 1, 2, $\infty$) \cite{KeeHY2016} and pyrochlore iridates \cite{Nagaosa2014} as potential topological systems \cite{Ueda2018, Guo2009, HuFiete2015,HuFietePRB2012}. The Ruddlesden-Popper series of iridates has a Mott metal-insulator transition, and the comparable magnitudes of the electron correlation and the spin-orbit coupling together with the crystal field interaction energies compete to elicit novel quantum states of matter including Dirac semimetals, Weyl semimetals \cite{Ueda2018}, topological Mott insulators and superconductors \cite{ZhangRev2018}. In spintronics the infinite member of the Ruddlesden-Popper series, perovskite \ce{SrIrO3}, is one of the most promising oxide materials for efficient spin-charge interconversion based on its large spin Hall angle \cite{Yoo2021, Fuchs2022} and spin orbit torque switching \cite{45_Yi_SOTswitching}, with applications in the generation, manipulation and detection of spin current for the development of robust, dissipationless, ultra fast computing devices \cite{44_Wang_oxidespintronics}.

Perovskite \ce{ABO3} has a simple cubic crystal structure with A cations in 12-fold coordination to the O ions, and B cations in octahedral coordination. The Goldschmidt tolerance factor determines the crystal structure type based on the ratio of the ionic radii of the A and B cations \cite{Goldschmidt}. The orthorhombic Pnma phase of perovskite typically forms because the A cations are too small relative to the B cations to fit into the B cation interstices. In such systems the \ce{BO6} octahedra are forced to follow a particular octahedral rotation pattern whose symmetry generates distinguishing half order reflections in reciprocal space. Perovskite \ce{SrIrO3} crystallises in the monoclinic C2/c structure in ambient conditions. The orthorhombic Pnma phase of \ce{SrIrO3} can only be synthesised in the bulk under high pressure (40 kbar) and high temperature (1000$^{\circ}$C) \cite{LONGO1971174} with a = 5.60 \AA, b = 7.89 \AA and c = 5.58 \AA, which inhibits the study of single crystals with this crystal structure. However, \ce{SrIrO3} can crystallise in the orthorhombic structure with the Pnma symmetry, illustrated in Fig. 1(a), in heterostructure geometry under epitaxial strain \cite{127_Guo}. The lattice mismatch between this structure and other perovskite single crystal substrates is known to introduce geometrical strain to minimise the interfacial energy in the growing film compressing the in-plane lattice parameters to match the dimensions of the substrate and extending the out of plane lattice parameter to preserve the volume of the unit cell. A recent report on \ce{SrIrO3} thin films determined lattice parameters of a = 5.6630 \AA, b = 7.9480 \AA and c = 5.6220 \AA{} in the Pnma space group setting \cite{KleinCrysEngComm} with a/b=0.7125, b/c=1.4137 and c/a=0.9928 giving a pseudocubic lattice parameter of a$_{pc}$ = 3.9819 \AA. 

In this system the formation of the orthorhombic phase can offer access to physical properties distinct from the monoclinic phase through modulation of the electronic bands at the Fermi energy. In the n = 1 member of the Ruddlesden-Popper series, Sr$_2$IrO$_4$, the Ir t$_{2g}$ bands from the octahedral crystal field are split by the spin-orbit interaction into narrow J$_{eff}$ = $\frac{1}{2}$ and two J$_{eff}$ = $\frac{3}{2}$ bands. The low correlation of the extended 5d orbitals is counteracted by the narrow J$_{eff}$ = $\frac{1}{2}$ bands making the correlation from the repulsive Hubbard interaction U, comparable to the spin-orbit interaction, and leading to a canted antiferromagnetic J$_{eff}$= $\frac{1}{2}$ Mott insulator \cite{Singh2015}. The octahedral connectivity strongly influences the electrical properties of the correlated Ruddlesden-Popper series \cite{206_Kawasaki_Schlom}. In the three-dimensional perovskite \ce{SrIrO3} the correlation energy and the increased hopping integral over the quasi two-dimensional n=1 member, which arises from the increased Ir-O-Ir interconnectivity between the layers along the c-axis, modulate the band structure resulting in a paramagnetic semimetal hosting Dirac nodal liens near the Fermi energy \cite {HYKee_DiracNodes2015}. The greater IrO$_6$ octahedral tilting with the tilt angle of 16$^{\circ}$ \cite{Zhao2008} varies the Ir-O-Ir bond angle and leads to the steric displacement of the Sr ions \cite{Singh2015} which influences the orbital overlap increasing the correlation energy in orthorhombic \ce{SrIrO3}. 

The growth of \ce{SrIrO3} has been increasingly studied in recent years with enhanced crystal growth control achieved in the 2-dimensional layer-by-layer growth of heterostructures on closely lattice matched perovskite single crystal substrates \cite{Groenendijk2016, 50_Jalan_MBE}. Such studies have enabled the close examination of the effect of epitaxial strain and relaxation of the crystal structure on the physical properties \cite{195_BISWAS2017}. The compressive strain-tuning of the semimetallic electrical properties of this correlated spin-orbit oxide has also been studied \cite{233_Chen, 35_Li} \cite{235_Gruenewald, 145_Pramanik} and the tuning of the spin-charge conversion efficiency with Ir stoichiometry \cite{11_Chen_Irstoich}. % The electronic band structure of the iridates has been studied by angle-resolved photoemission spectroscopy \cite{77_Dhingra}.The optical properties have been researched by Liu et al. \cite{259_Liu_Opt}. 
This system has also been widely studied with theoretical works reportedly predicting a 2-dimensional topological insulator with band inversion arising from the spin orbit coupling in the honeycomb lattice of a perovskite bilayer grown along the (111) axis \cite{Xiao2011}. Calculations performed by Zeb et al. uncovered Dirac nodal points close to the Fermi level in the perovskite \cite{ZebKee2012}. One first-principles study of strain-induced octahedral tilting observed tunable transport properties including a computed insulating state \cite{Singh2015}. %The effect of the substrate on optical properties and the electronic and magnetic properties have been studied by density functional theory with careful selection of Coulomb correlation U \cite{215_Min, 198_Franchini}. This material has also been studied for it potential applications as an oxygen evolution electrocatalyst \cite{31_Zhang, 23_Zhang, 14_Wang} in addition to its topological and spintronic properties including anomalous Hall effect due to proximity effect \cite{70_Yoo} and spin-charge conversion efficiency \cite{9_Fuchs, 11_Chen_Irstoich}.

%Manipulating the growth of atomic planes along particular crystallographic axes, such as the (001) and (111), within the perovskite crystal lattice has the potential to enable access to structural motifs with the broken spatial symmetries to realise topological phases such as buckled honeycomb lattice \cite{liu_chakhalian_2016} analogous to the Haldane model of graphene which predicted quantum Hall effect without Landau levels \cite{Haldane1988}. Advances in the crystalline quality and unit cell layered growth control of oxides in recent years using techniques such as oxide molecular beam epitaxy (MBE) and pulsed laser deposition (PLD) has made it possible to approach the degree of crystalline perfection and thickness control required to replicate the theoretical models. The complex interactions at the interface in this heavy transition metal oxide can potentially be modified by strain engineering and symmetry breaking to yield novel interface properties in the vicinity of the metal-insulator phase boundary. It is important to understand the ways in which we can tune these properties to enhance functionality in this promising material. We therefore studied the influence of the deposition partial pressure on the structural, electrical transport, optical and electronic properties of this system as it relates to oxygen stoichiometry and compared these properties with the density functional theory computed properties.

\section{Methods}
\subsection{Experimental methods}
Heterostructure growth was carried out by pulsed laser deposition from a polycrystalline target. Monoclinic \ce{SrIrO3} source material was synthesised by solid state reaction according to a previously reported method \cite{Groenendijk2016} where stoichiometric quantities of \ce{SrCO3} and \ce{IrO2} were mixed, ground and heated to 950$^{\circ}$C for 12 hours in air before being ground and heated to 1050$^{\circ}$C for 24 hours. The monoclinic phase of the resultant powder was confirmed by powder X-ray diffraction (Rigaku Smartlab XRD, Cu $K_{\alpha 1}$ radiation) A dense single-phase target was subsequently prepared by spark plasma sintering (SPS) \cite{TianSPS} of the powder heating to 950$^{\circ}$C for 2 mins in He under a pressure of 60 MPa. Thin films were grown on low miscut (001) oriented \ce{SrTiO3} single crystal wafers from PiKem. Atomically flat (001) \ce{SrTiO3} surfaces were prepared by etching in buffered HF acid for 30 s with prior sonication in acetone, ethanol and deionised water for 10 mins each following the procedure of Biswas et al. \cite{Biswas_HFetch}. A subsequent annealing at 1050$^{\circ}$C for 2 hours in air resulted in a step-and-terrace termination which was observed by atomic force microscopy with steps corresponding to one unit cell in height. The lattice mismatch between \ce{SrIrO3} and the \ce{SrTiO3} substrate is $\Delta$$\epsilon$ =  -2.0\% which introduces compressive strain in the interfacial layers. The deposition was carried out in a Neocera PLD system with a base pressure of $1\cdot 10^{-8}$ Torr using a KrF excimer laser ($\lambda$ = 248 nm) with pulse repetition rates varying from 2 - 10 Hz and laser fluences in the range 1.9 - 3.6 J/cm$^2$.  The deposition growth window was explored, with temperatures in the range of 600-700$^{\circ}$C and oxygen partial pressures of 20-70 mTorr, to examine the effect of subtle changes in growth parameters on the physical and structural properties of \ce{SrIrO3} thin films on (001) \ce{SrTiO3} single crystal substrates. After deposition the samples were cooled in the deposition partial pressure of oxygen at a rate of 10 $^{\circ}$C/minute. The phase, orientation, out-of-plane lattice spacing, crystallinity and mosaicity were examined using a high-resolution (0.0002$^{\circ}$) four-circle diffractometer (9 kW rotating anode Cu $K_{\alpha 1}$ radiation with wavelength 1.540593 \AA{}). The scans were collected in the parallel beam geometry with a 2 bounce Ge (220) monochromator using a HyPix-3000 2D detector in 1D mode. The thickness of the samples was measured using X-ray reflectometry and the in-plane lattice parameters were determined using asymmetric reciprocal space mapping. The electrical transport properties were measured in the Van der Pauw configuration \cite{VanderPauw1958} with a 14 T Quantum Design Dynacool ETO using In solder and Au wires and an AC excitation with an amplitude of 0.5 $\mu$A and a frequency of 70.19043 Hz. The antisymmetric component of the transverse Hall magnetoresistance was used to calculate the effective carrier concentration and mobility of the films. The electronic structure of the heterostructures was probed by X-ray photoemission spectroscopy (XPS) using a monochromated Al $K_{\alpha}$ photon source (h$\nu$ = 1486.6 eV) under ultra-high vacuum ($3\cdot 10^{-11}$ mbar) to investigate the valence band and semicore level energies at the \ce{SrIrO3} film surface. Emitted photoelectrons were detected using an Omicron SPHERA electrostatic hemispherical deflection analyser (mean radius 125mm) and the valence band, Sr 3$p$ and Ir 4$d$, Ir 4$f$, O 1$s$ and Sr 3$d$ level spectra were collected with an instrumental energy resolution of 0.4 eV. The binding energy was calibrated using a sputtered polycrystalline metallic silver reference sample. The film thicknesses were determined by spectroscopic ellipsometry using a Woollam M200UI ellipsometer, with wavelength range 240–1700 nm and photon energies between 0.73–5.14 eV, at three different angles (65$^\circ$, 70$^\circ$, and 75$^\circ$).

\subsection{Computational methods}
Density functional theory calculations were carried out within the generalized gradient approximation (GGA) using the Vienna Ab initio Simulation Package \cite{VASP} in the plane-wave basis of the projector augmented wave method \cite{PAW}. The GGA+U calculations \cite{AnisimovGGAU} were carried out including the correlation effects of Ir in the Hubbard U in accordance with a study of Manca et al. \cite{MancaN_GGAU} which determined optimum values of U=0.8 eV and J=0.12 eV for a nonmagnetic bulk compound and the Perdew-Burke-Ernzerhof \cite{PBE} functional was used to correct the exchange correlation with a plane-wave cutoff of 520 eV. The spin-orbit interaction was included in all calculations with a gamma-centered 12$\times$12$\times$9 k-point grid. Galore software package \cite{Galore2018} was used to convolve the partial densities of states for comparison of the valence edge of the samples assuming a Gaussian instrumental broadening of 0.4 eV and Lorentzian lifetime broadening of 0.5 eV.
%Structural relaxations were carried out with the atomic positions and lattice constants allowed to relax until calculations converged when the forces on all atoms were less than 0.01 eV/\AA. 

\section{Results and discussion}
We have studied the sensitive response of the growth regime and phase diagram of \ce{SrIrO3} heterostructures on \ce{SrTiO3} substrates grown by pulsed laser deposition to subtle changes in growth parameters. A range of substrate temperatures, oxygen partial pressures, laser fluence and repetition rates and film thicknesses were explored to optimise the growth of this heterosystem. While higher temperatures are frequently used in PLD, this was avoided due to the high vapour pressure of the Ir component in order to achieve stoichiometric transfer of both cations of the source composition. Temperatures in the range of 600-700\textcelsius{} were therefore explored, yielding highly crystalline epitaxial films at 675\textcelsius{}. Relatively high oxygen partial pressures were also employed as this system has a narrow range of pressure for which an appreciable growth rate can be achieved. A sharp change in growth regime of \ce{SrIrO3} was indeed observed over a narrow range of oxygen partial pressure, and the growth rate increased rapidly with the slightest increase in pressure yielding a thick film, whose consequent structural, electrical, optical and electronic properties are markedly different from the lower pressure sample. The electrical properties of this semimetallic perovskite are reported to vary with Ir stoichiometry \cite{11_Chen_Irstoich} and strain \cite{233_Chen} and we provide evidence for the tuning of the majority carrier type in this correlated system through deposition partial pressure and geometric strain. A comparison study of the structural and physical properties of samples resulting from two sets of growth conditions is detailed in the following sections.

\begin{figure*}
        \centering
        \includegraphics[width=\columnwidth]{"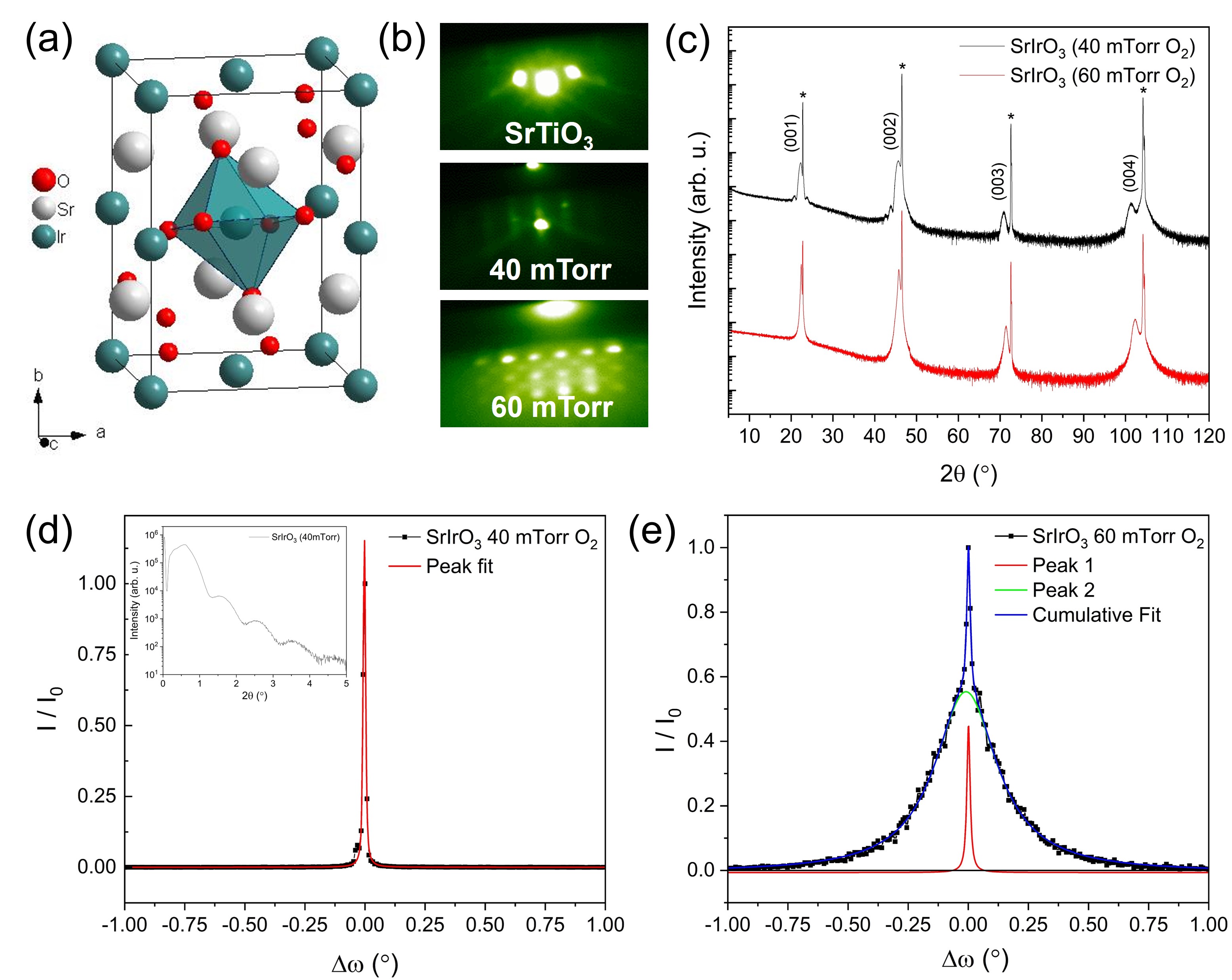"}
        \caption{(a) \ce{SrIrO3} crystal structure in the Pnma space group setting. (b) RHEED patterns of the STO substrate prior to deposition (top) and after deposition for the 40 mTorr (middle) sample with rods indicating 2-dimensional growth, and for the 60 mTorr (bottom) sample with spots in the diffraction pattern indicating 3-dimensional island growth post-deposition. (c) 2$\theta$ X-ray diffraction scan of (00$l$)$_{pc}$ oriented \ce{SrIrO3}/\ce{SrTiO3} heterostructures deposited at (i) 40 mTorr (black) and (ii) 60 mTorr (red). (d) XRD $\omega$ scan showing rocking curve of the 40 mTorr sample which was fitted to a pseudo-Voigt peak, with a narrow full width at half maximum height indicating superior crystalline quality. Inset shows X-ray reflectivity scan of this sample which was fitted to determine the film thickness. (e) Rocking curve of the 60 mTorr sample which was fitted to two pseudo-Voigt functions to extract a broad basal peak and a sharp peak at high intensity.}
    \end{figure*}

\subsection{Heterostructure growth and structural properties}
The case study samples were both deposited at a substrate temperature of 675\textcelsius{} with a laser fluence of 3.54 J/cm$^2$, a laser repetition rate of 5Hz and 50000 laser pulses, however different oxygen partial pressures of 40 mTorr and 60 mTorr were examined. The growth of these heterostructures was monitored by reflection high-energy electron diffraction (RHEED) in real-time during the depositions, on the prepared \ce{SrTiO3} surface as shown in Fig. 1(b, top). This indicated that the initial growth regime was 2-dimensional for both films, as demonstrated by the rods in the electron diffraction pattern in Fig. 1(b, middle) for the sample grown at 40 mTorr. 2-dimensional growth persisted up to a certain thickness before 3-dimensional island growth was later assumed as shown in Fig. 1(b, bottom). Both sets of conditions yielded highly crystalline epitaxial films corresponding to the orthorhombic phase of perovskite \ce{SrIrO3} which were oriented along the pseudocubic (00$l$) axis as indicated by X-ray diffraction patterns in Fig. 1(c). Pendellosung fringes about the Bragg reflections indicated a high coherence of growth in the thinner sample grown at 40 mTorr. The X-ray reflectivity was used to determine the thickness and surface and interfacial roughness of the films, as shown in Fig. 1(d) inset. At the slightly lower pressure, 40 mTorr, the growth rate was laboured yielding a thin film of 8.5 nm thickness, and sustaining a 2-dimensional growth regime. While at slightly higher pressure, 60 mTorr, the marginal change in this parameter resulted in a film which was at least one order of magnitude thicker than the previous sample with 3-dimensional growth. The sample grown at 60 mTorr was too thick to determine the thickness using this technique and was determined by ellipsometry to be 92 nm. The rocking curve, shown in Fig. 1(d), indicated an extremely sharp full width at half maximum height peak for the 40 mTorr deposited sample which was 0.0124(1)$^{\circ}$, the order of the width of the substrate rocking curve, demonstrating the extremely high crystalline quality and low mosaicity of the sample. The higher pressure growth resulted in a rocking curve with a broad basal peak together with a sharp high intensity narrow peak as shown in Fig. 1(e). The cumulative rocking curve was fitted to two Gaussian peaks with widths of 0.0217(8)$^{\circ}$ and 0.327(3)$^{\circ}$. The broad lower intensity peak may be due to the partial strain of the film close to the interface which relaxed giving rise to the narrow peak from the highly crystalline continuing film.

\begin{figure*}
    \centering
    \includegraphics[width=\columnwidth]{"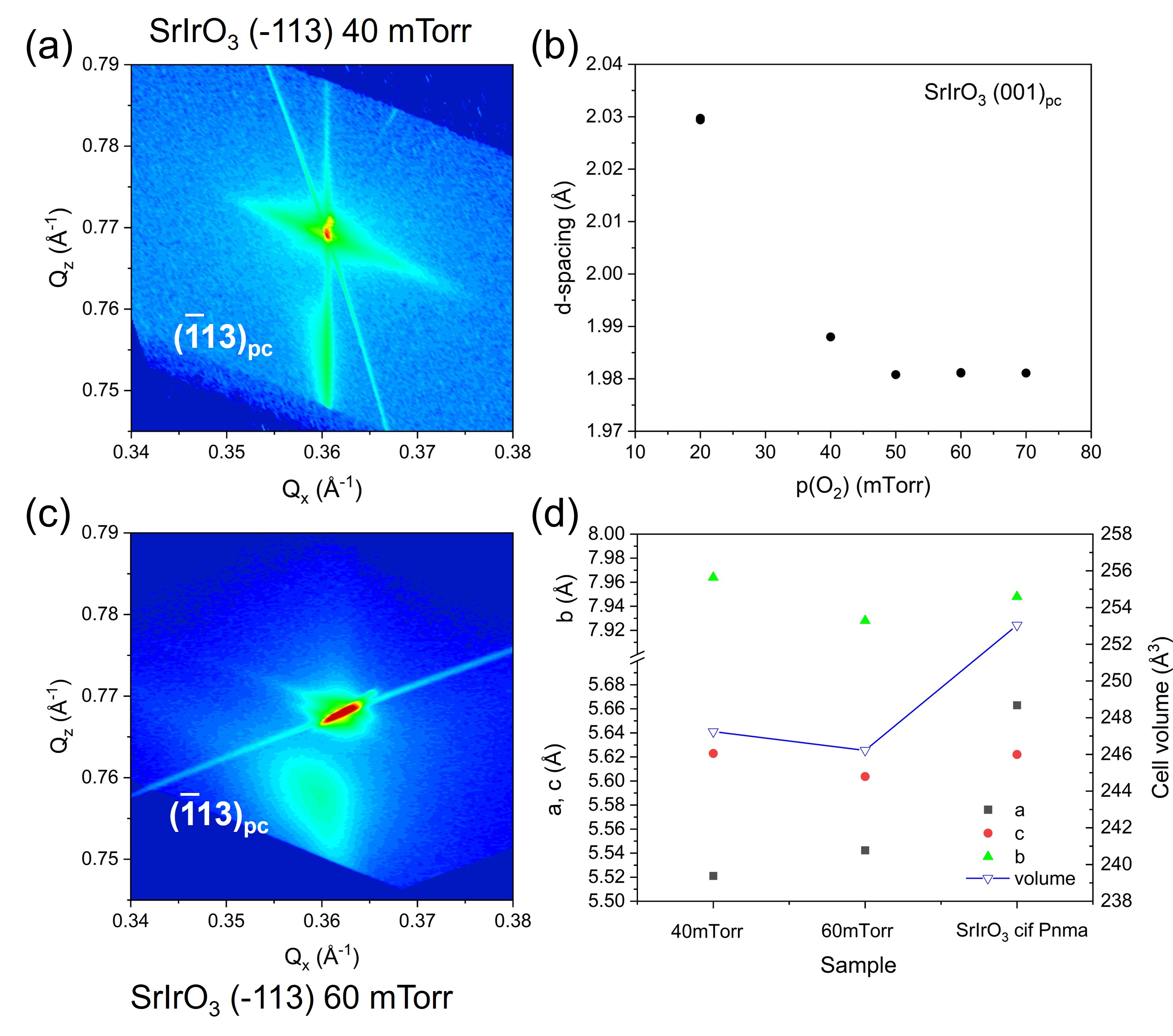"}
    \caption{(a)Reciprocal space map of the \ce{SrIrO3}/\ce{SrTiO3} heterostructure sample deposited at 40 mTorr of O$_2$ capturing the ($\bar{1}13$) interplanar reflection of the film which is aligned to the substrate in $Q_X$ indicating that the sample is geometrically strained to the STO substrate in-plane. (b) Out-of-plane d-spacings of the (001) reflections of the thin films deposited at oxygen partial pressures between 20 and 70 mTorr. (c) Reciprocal space map of the heterostructure grown at 60 mTorr oxygen partial pressure p(O$_2$) reveals a broader mosaicity of the film reflection which is slightly misaligned from the ($\bar{1}13$) reflection of the substrate indicating that the lattice parameters of this film are relaxed. (d) Extracted a, b and c parameters and unit cell volume (blue) of the orthorhombic phase of \ce{SrIrO3} for either case compared with the reported bulk parameters in the literature.}
\end{figure*}

Reciprocal space maps demonstrated the alignment of the pseudocubic (013) and ($\bar{1}13$) reflections of the thinner sample with the cubic analogues of the substrate along $Q_x$ indicating that it was strained to the STO substrate with a narrow profile as shown in Fig. 2(a). The out-of-plane d-spacing was observed by XRD to decrease with increasing oxygen pressure during growth from 2.0297\AA{} to 1.9811\AA{} as shown in Fig. 2(b) for each of the sampled deposition pressures between 20 and 70 mTorr. Values of d-spacing of 1.9880\AA{} and 1.9812\AA{} were recorded for the (001) reflections of the films grown at 40 mTorr and 60 mTorr respectively. Similar reciprocal space map analysis of the ($\bar{1}13$) pseudocubic reflection of the thicker film demonstrated a broader mosaicity of the film peak which was broadly misaligned with the substrate in Fig. 2(c) indicating that although closely lattice matched this sample was semi-relaxed. The lattice parameters were determined from the reciprocal space maps to be a = 5.52109\AA{}, b = 7.964\AA{}, c = 5.62291\AA{} for the thin sample and a = 5.5423\AA{}, b = 7.928\AA{}, c = 5.60368\AA{} for the thick sample. The lattice parameters for the reported thin film sample for comparison were a = 5.6630\AA{}, b = 7.9480\AA{} and c = 5.6220\AA{} as shown in Fig. 2(d). In both films the a parameter was significantly reduced while the b and c parameters were only slightly varied. This contraction of the a parameter is likely due to the compressive epitaxial strain from the STO substrate. This is also reflected in the unchanged b/c ratios between the reported lattice parameters, b/c = 1.4137, and both case study samples, b/c = 1.4163 for the 40 mTorr and 1.4148 for the 60 mTorr samples. The c/a ratio was increased however from 0.9928 to 1.0184 for the 40 mTorr film and 1.0111 for the 60 mTorr sample. The a/b ratio was observed to be reduced in both cases from 0.7125 to 0.6933 for the 40 mTorr film and 0.6991 for the 60 mTorr film, indicating a greater degree of tetragonality. The cell volume was slightly reduced in both films compared with the reported volume as shown in Fig. 2(d).
%a/b = 0.6933, b/c = 1.4163, c/a = 1.0184 (060323SIO). a/b = 0.6991, b/c = 1.4148, c/a = 1.0111 (170323SIO).
%The film grown at 40 mTorr had an out-of-plane lattice parameter of 7.964\AA{}, while the 60 mTorr sample had an out-of-plane lattice constant of 7.928\AA{}. These figures were notably different from the reported values which varied from 7.948\AA{} (cif 196446) to 7.8821\AA{} (cif 1627982) demonstrating how sensitive this system is to stoichiometry of the composition. 

\begin{figure*}
    \centering
    \includegraphics[width=\columnwidth]{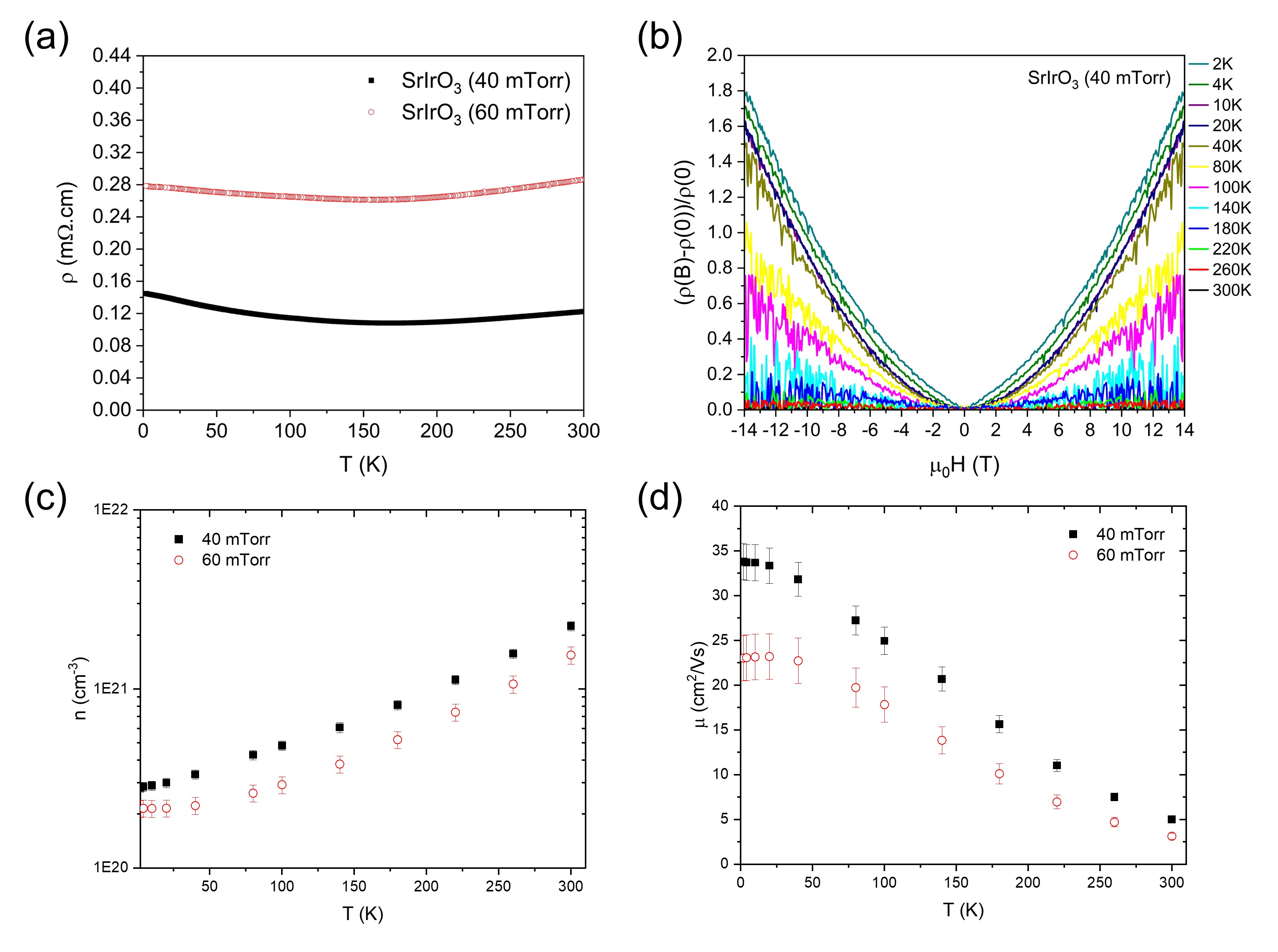}
    \caption{(a) The temperature dependence of the electrical resistivity of \ce{SrIrO3}/\ce{SrTiO3} heterostructures grown at 40 mTorr (black) and 60 mTorr O$_2$ (red). (b) The magnetoresistance of the sample grown at 40 mTorr in magnetic fields of up to 14T in the temperature range 2-300K. (c) Effective carrier concentrations and (d) mobilities of the sample grown at 40 mTorr (black squares) and 60 mTorr (red circles) in field sweeps up to 14T at various temperatures extracted from Hall effect measurements.}
\end{figure*}

%Both film orientations show a similar metallic temperature dependence in the electrical resistivity with a small negative temperature coefficient across the measured temperature range (Fig. 4). This cannot be explained by a thermally activated semiconducting mechanism and is similar to previously reported thin film \cite{Nakajima2014,Chiba2019471} and polycrystalline samples prepared in air \cite{Carbonio1999361}. Resistivity values of $\rho$ = 0.54(3) m$\Omega$cm and $\rho$ = 0.43(3) m$\Omega$cm were obtained for the (001) and the (111) oriented layers respectively at 300 K, in close agreement with the values reported in polycrystalline samples \cite{Kanno1993106,Carbonio1999361}, single crystals \cite{PhysRevB.73.193107} and thin films \cite{Nakajima2014,Chiba2019471}. The reduced resistivity observed in the (111) oriented film could be attributed to lower scattering from dislocations as the crystallinity of this film is superior. 

\subsection{Electrical properties}
The two case samples considered  differ in degree of strain and thickness due to the slight variation in oxygen deposition pressure (40 mTorr and 60 mTorr), they were therefore selected for closer inspection of the electrical resistivity and magnetotransport in order to gauge the extent to which the electrical properties have been influenced by these differences and determine whether the oxygen stoichiometry is responsible for these incongruences. Both heterostructure samples exhibited a metallic temperature dependence, as shown in Fig. 3(a), which did not follow any standard power law, consistent with the expected level of correlation in the system. A change in temperature coefficient from positive to negative was observed upon cooling through 170K in the measured temperature range. The up-turn in the resistivity is not as sharp as observed in thin \ce{SrIrO3} films where weak localization is observed\cite{11_Chen_Irstoich} \cite{BiswasJAP2014_MIT}. Resistivity values of $\rho$ = 0.122492 m$\Omega$cm at 300K and $\rho$ = 0.144787 m$\Omega$cm at 2K were observed for the low-pressure (40 mTorr) sample, which are an order of magnitude lower than reported thin film values grown on \ce{DyScO3} \cite{11_Chen_Irstoich} and on \ce{GdScO3} \cite{BiswasJAP2014_MIT}. This may indicate that oxygen deficiency due to the slightly more reduced atmosphere during growth may also contribute carriers to enhance the conductivity or that nonstoichiometric Ir deficiency may be induced by the reduced oxygen partial pressure as Yi et al. have inferred \cite{11_Chen_Irstoich}. We have observed that the resistivity increases for the higher pressure (60 mTorr) sample while Yi et al. observed a resistivity minimum for 75 mTorr p(O$_2$).

The symmetric component of the magnetotransport indicated a positive open orbit magnetoconductance with a parabolic profile is observed down to 2K for both samples with a amgnitude of around 2\% in the film grown at 40 mTorr as shown in Fig. 3(b). The antisymmetric component of the field-dependence of the resistance was linear for both samples with a positive slope down to 2K indicating a single carrier type is responsible for the conductivity. The extracted transport parameters indicated similar temperature dependence of the carrier concentrations and mobilities for both samples (Fig. 3(c) and (d)). The carrier concentration is slightly larger in the sample grown at 40 mTorr (2.2(1)$\cdot 10^{21}$ cm$^{-3}$ at room temperature and 2.8(2)$\cdot 10^{20}$ cm$^{-3}$ at 2K) than in the sample grown at 60 mTorr (1.5(1)$\cdot 10^{21}$ cm$^{-3}$ at room temperature and 2.1(2)$\cdot 10^{20}$ cm$^{-3}$ at 2K) and it increases by an order of magnitude going from 2K to room temperature.%. The carrier concentrations decreased monotonically with decreasing temperature indicating a degree of thermal activation of the Fermi distribution at higher temperatures. 
The carrier concentration is similar to the reported carrier concentrations of 4.6 $\cdot 10^{20}$ cm$^{-3}$ for \ce{SrIrO3} heterostructures grown on \ce{SrTiO3} at 75 mTorr \cite{Fuchs2022} and it is less than the theoretical carrier concentration of 1.58$\cdot 10^{22}$ cm$^{-3}$ for 4 electrons per Ir. This may indicate that some correlation effects are limiting the participation of some of the carriers in the conduction mechanism and that the system is close to the metal-insulator phase boundary. The temperature dependence of the mobilities follow similar behaviour for both samples and is typical of a phonon scattering limited mechanism at high temperature with carrier mobility of about 5 cm$^2$/Vs at ambient temperature for both samples, increasing to a saturation value of 33(2) and 23(3) cm$^2$/Vs at 2K for the sample grown at 40 and 60 mTorr respectively. %Again, this deviates from reported mobilities of XXXX \cite{} which may be attributed to an increased effective mass or increased scattering time of the carriers. The reduced carrier mobility at higher temperatures indicates the dominant scattering mechanism at high temperature is phonon scattering. 

%The transport parameters for the thick sample (60 mTorr 170323SIO) indicated… 
%\begin{figure*}
%    \centering
%    \includegraphics[width=\columnwidth]{"Fig4_OpticalCond_Ellipsometry_NEDOS10000_1.jpg"}
%    \caption{(a) Spectroscopic ellipsometry measurement of the complex dielectric function of the \ce{SrIrO3} films grown at (a) 40 mTorr and (b) 60 mTorr. (c) The extracted optical conductivity of both samples, showing the lower-pressure sample (black) and the higher-pressure sample (red), together with the optical conductivity of the calculated orthorhombic \ce{SrIrO3} structure from the DFT+U calculations (blue).}
%\end{figure*}

%\subsection{Optical properties}
%Ellipsometry measurements were carried out on each of the samples to determine both the thickness of the high-pressure sample, the optical conductivity and the optical transmittance in the UV and visible ranges of the electromagnetic spectrum.

\begin{figure*}
    \centering
    \includegraphics[width=\columnwidth]{"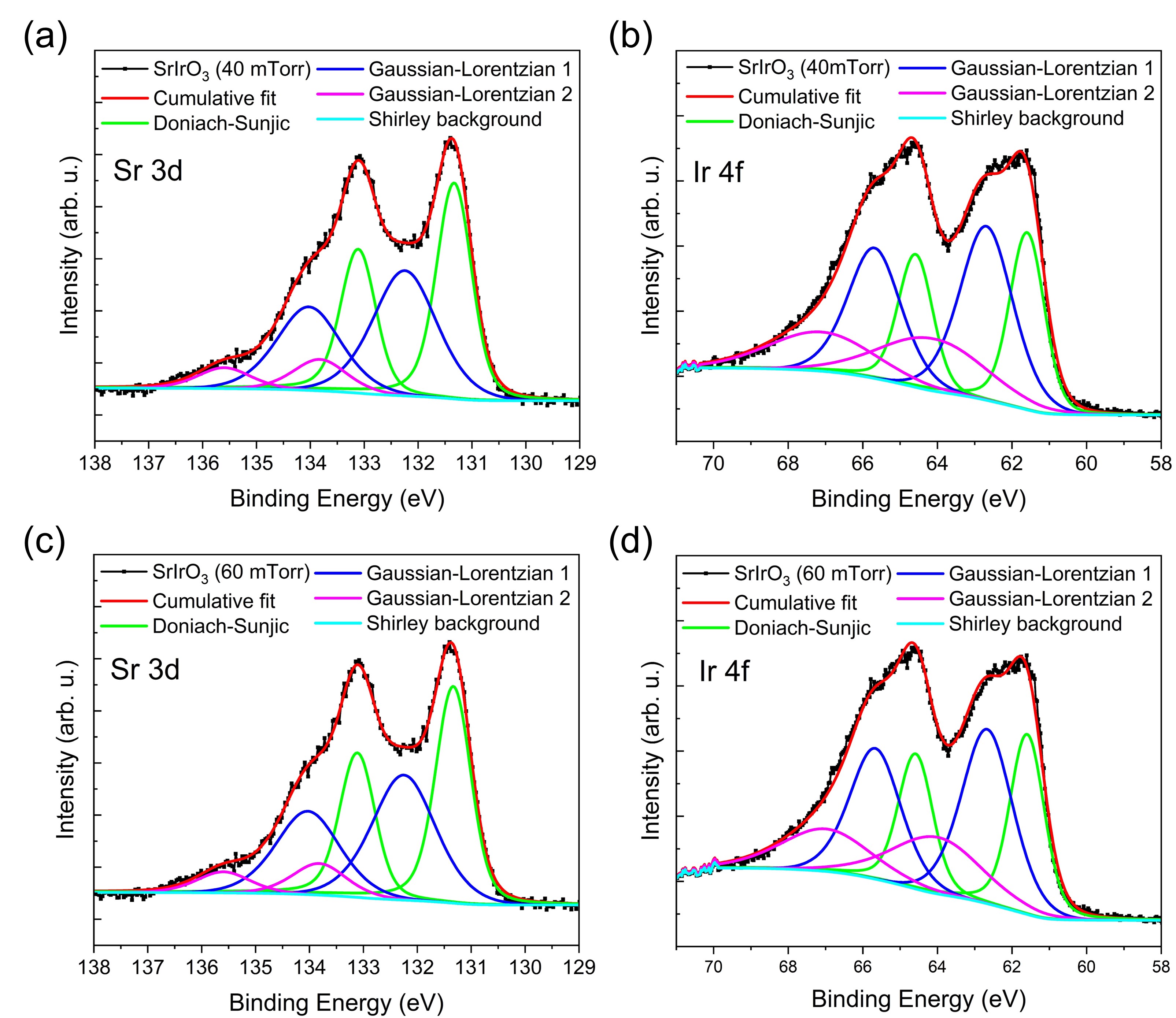"}
    \caption{X-ray photoemission spectra for the sample grown at 40 mTorr showing (a) the Sr 3$d$ core lines and (b) the peaks corresponding to the Ir 4$f$ lines. Spectral lines corresponding to (c) the Sr 3$d$ levels and (d) the Ir 4$f$ for the sample grown at 60 mTorr.}
\end{figure*}

\subsection{Electronic Structure and Density Functional Theory Calculations}
X-ray photoemission spectra were collected for both the 40 mTorr and 60 mTorr samples of \ce{SrIrO3} to probe the oxidation states and the local chemical environment of the elements including the core Sr 3$p$, Ir 4$d$, Ir 4$f$, O 1$s$ and Sr 3$d$ levels together with the valence band region. The Sr 3$p$ and Ir 4$d$ core levels in both samples (shown in Fig. 4(a-d)) with Shirley-backgrounds subtracted had symmetric profiles and were therefore fitted with single Gaussian-Lorentzian doublets corresponding to the Sr 3$p\frac{1}{2}$ and 3$p\frac{3}{2}$, and Ir 4$d\frac{5}{2}$ and 4$d\frac{3}{2}$ lines/transitions with little difference observed between the two samples. The Sr 3d spectral peaks were broader with asymmetric profiles and they were fitted with three doublets accounting for the Sr 3$d\frac{5}{2}$ and 3$d\frac{3}{2}$ levels in both samples. Metallic systems are known to exhibit asymmetric peak profiles due to scattering of core-hole potential \cite{DemkovXPS}. The broadening of the peaks may indicate a change in the number of chemical bonds to the Sr ion contributing to the peak shape. The multiple doublets in this metallic phase may indicate differences in the chemical environment experienced by this element corresponding to localised correlated orbitals and uncorrelated orbitals, or possibly due to different oxidation states at the surface and in the bulk of the film. Similarly, the Ir 4$f$ spectral lines with backgrounds subtracted were fitted with three doublets using a Marquardt fitting method/algorithm to fit the Ir 4$f\frac{7}{2}$ and 4$f\frac{5}{2}$ peaks. The fit returns a narrow Gaussian-Lorentzian doublet with energy positions close to those reported for metallic Ir, a broader Gaussian-Lorentzian doublet with positions corresponding to reports for \ce{IrO2}, and a further broad low area/intensity doublet pair at higher binding energy. A shift of ~5 meV of the Ir 4$f$ peaks to slightly higher binding energy is observed in the sample grown at 60 mTorr relative to the sample grown at 40 mTorr which is consistent with the transport data.

%O 1s

DFT calculations were performed for orthorhombic \ce{SrIrO3} to obtain the partial density of states (PDOS) in the valence-band level shown in Fig. 5(a). The valence band is dominated by the O 2$p$ and Ir 5$d$ orbitals. The binding energy spectra of electrons below the VBM were acquired and the Fermi level was set to the zero of the binding energy-scale. The one-electron angular-corrected photoionisation cross sections for the X-ray energy $h\nu$ = 1486.6eV have been applied to the calculated PDOS using Galore software \cite{Galore2018}. The weighted densities of states including the O 2$p$ and Ir 5$d$ orbitals were convoluted with 0.4 eV Gaussian instrumental broadening and 0.5 eV Lorentzian lifetime broadening. The photoemission spectra have been corrected according to the calibrations and the calculated DFT data have been shifted to fit the spectra, corrected for the Shirley-background, as the DFT convention positions the zero-energy point at the valence band maximum. Comparison of the DFT calculations with the Fermi edge of the XPS spectra indicate that they map well to the spectra for the LDA+U calculations with U = 0.8. In transition metal oxides the valence states include oxygen 2$p$ and the transition metal $d$ orbitals which we observe in the Ir 5$d$ states (Fig. 5(a)). The Ir$^{4+}$ ions in \ce{SrIrO3} have a nominal $d^5$ electron configuration of the Ir 5$d$ orbital. The spectral intensity in the vicinity of the valence edge can be largely accounted for by the O 2$p\frac{1}{2}$ and 2$p\frac{3}{2}$ bands together with the Ir 5$d\frac{5}{2}$ and 5$d\frac{3}{2}$ levels. 

A shift of ~ 0.3 eV is observed in the valence band edge between the two samples, with the intensity dropping close to zero in the sample grown at 60 mTorr while the zero energy cuts through the intensity of the sample grown at 40 mTorr (Fig. 5(b)). This could be interpreted as the higher-pressure growth resulted in an electronic structure at the Fermi level close to a semimetallic character while the reduced pressure yielded a p-type metallic sample.

The secondary electron cut-off energy of the \ce{SrIrO3} film was measured with a bias applied to the sample to avoid overlap with the analyzer. The work function of the (001)$_{pc}$ surface of \ce{SrIrO3} was determined from this energy cut-off to be $\phi$ = 5.3(2) eV.

\begin{figure*}
    \centering
    \includegraphics[width=\columnwidth]{"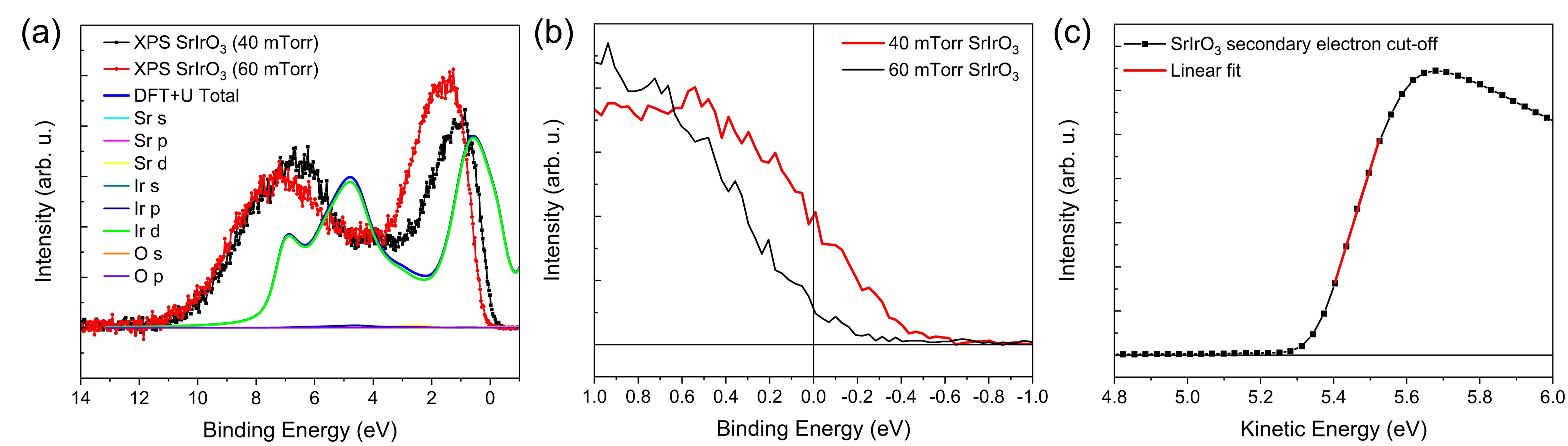"}
    \caption{(a) XPS spectra corresponding to the occupied bands close to the Fermi edge for the \ce{SrIrO3} samples grown at 40 mTorr (red) and 60 mTorr (black) together with the density of states convolved with instrumental and lifetime broadening from the DFT+U calculations, comprised mainly by Ir $d$ states. (b) Zoom on the Fermi edge of both films indicating a shift in the number of contributing states contributing to the spectral intensity at the Fermi level. (c) Secondary electron cut-off energy spectrum of \ce{SrIrO3} film as a function of kinetic energy.}
\end{figure*}

\section{Conclusions}
The growth of \ce{SrIrO3} heterostructures is highly sensitive to the deposition conditions. The growth rate is observed to increase dramatically with a slight increase in oxygen partial pressure. The films obtained at lower pressure had a superior crystallinity and the growth mode was 2-dimensional. Increasing the pressure was noted to reduce the crystallinity of the layer with rocking curves composed of two overlapping peaks, one broad basal peak and a second narrow high intensity peak, it also changed the growth regime to 3-dimensional growth. Reciprocal space maps indicate the film grown at 40 mTorr is compressively strained to the substrate while the sample at 60 mTorr is relaxed while closely lattice matched. The higher oxygen pressure is observed to decrease the out-of-plane d-spacing. The combined effects of compressive strain and lower oxygen partial pressure are observed to significantly reduce the a lattice constant, while moderately increasing the b parameter and only slightly reducing the c parameter in comparison to the bulk introducing a lattice distortion. Higher growth pressure in the relaxed film is noted to considerably reduce a and b, while slightly decreasing c, indicating the volume of the unit cell is changing and that there may be a structural distortion.
The electronic properties and electronic structures varied slightly with the growth pressure and the careful examination of the transport properties such as carrier concentration and valence band edge shift allowed to postulate that it is possible to tune the Fermi surface of this correlated metallic oxide from a semimetal to a p-type metal. These observations together with the growth control achieved in this study will be useful to design heterostructure based on this compound and further physical characterisation such as optical properties can further our understanding of correlated oxides.

\nocite{*}

\section*{Acknowledgements}
The authors would like to acknowledge financial support for MOS from the Engineering and Physical Sciences Research Council (EPSRC) as part of a Daphne Jackson Fellowship. The X-ray diffraction facility used to characterize the films was supported by EPSRC (grant EP/P001513/1). We thank the Leverhulme Trust for the use of facilities in the Leverhulme Research Centre for Functional Materials Design. 

\section*{Authors declaration}
\subsection*{Conflict of interest}
The authors have no conflicts to disclose.
\subsection*{Author contributions}
Tim Veal: Investigation (XPS measurements and data analysis); Writing - Review \& Editing. Wesley Surta: Spark plasma sintering of \ce{SrIrO3} source target. Marita O'Sullivan: Conceptualization; Formal analysis; Funding acquisition; Investigation (Heterostructure growth, XRD, transport, ellipsometry, DFT); Methodology; Project administration; Resources; Validation; Visualization; Writing - original draft, review \& editing.

\section*{Data availability}
The data that support the findings of this study are openly available in 
\section*{References}
\bibliography{aipsamp}% Produces the bibliography via BibTeX.

\end{document}